\documentstyle[preprint,aps,epsfig,eqsecnum,tighten,floats]{revtex}

\begin{document}
\draft
\title{Absence of self-averaging in the complex admittance \\
for transport through random media\\
}
\author{Mitsuhiro Kawasaki and Takashi Odagaki\\
}
\address{
Department of Physics, Kyushu University, Fukuoka 812-8581, Japan
}
\author{Klaus W. Kehr}
\address{
Institut f\"{u}r Festk\"{o}rperforschung, Forschungszentrum J\"{u}lich GmbH,
D-52425 J\"{u}lich, Germany}
\date{\today}
\maketitle
\begin{abstract}
A random walk model in a one dimensional disordered medium with 
an oscillatory input current is presented as a generic model of 
boundary perturbation methods to investigate properties of a 
transport process in a disordered medium.
It is rigorously shown that an admittance which is equal to the 
Fourier-Laplace transform of the first-passage time distribution is 
non-self-averaging when the disorder is strong.
The low frequency behavior of the disorder-averaged admittance, 
$\langle \chi \rangle -1 \sim \omega^{\mu}$ where $\mu < 1$, 
does not coincide with 
the low frequency behavior of the admittance for any sample, 
$\chi - 1 \sim \omega$. 
It implies that the Cole-Cole plot of $\langle \chi \rangle$ appears 
at a different position from the Cole-Cole plots of $\chi$ of any sample. 
These results are confirmed by Monte-Carlo simulations.
\end{abstract}
\pacs{PACS numbers: 02.50.Ey, 05.40.Fb, 72.80.Ng}

\narrowtext

Response of a system to an external field is a standard information 
to be utilized in the study of condensed matter. Recently, 
the frequency response method 
\cite{Yasuda95} and the intensity modulated photocurrent spectroscopy 
\cite{Jongh96} have been introduced, where an oscillatory 
perturbation is applied at one end of a system and a response 
from the other end of the system is measured. 
Thus, these experimental techniques can be considered to belong to a 
generic method which can be called the boundary perturbation method (BPM). 
In the presence of a periodically forced boundary condition on one end of 
a system, the output from the other end of the system is in proportion 
to the perturbation in the linear regime and the proportionality constant is 
called the admittance. The frequency dependence of the admittance contains 
various information of the dynamics of the system.
The BPM is expected to provide useful information on transport properties of 
disordered media.

The analysis of transport of particles in a random medium attracts 
great interest since the transport mechanism is basic for the 
understanding of many physical phenomena, from electrical 
conductivity to thermal properties \cite{general interest}. 
Usually it is assumed that a sample used in experiments is sufficiently large 
that it can be considered to be composed of a large number of sub-systems.
If each of sub-systems is itself macroscopic, 
the boundary effect is negligible and each sub-system may be considered
to be a realization of the system with a particular choice for the disorder. 
Thus, a measurement of any observable in such a system corresponds to 
an average over all the sub-systems, i.e. an average over the ensemble of 
all realizations of the disorder. Systems for which this assumption is valid 
are said to be self-averaging (SA). Since the assumption implies 
no sample dependence, 
it guarantees reproducibility of experimental results in any sample.

However, it is known that the disorder-averaged quantity follows a dynamics 
that is different from the dynamics of the quantity of each sample.
If the dynamics of particles in a disordered sample is assumed 
to be stochastic and Markovian, i.e. to follow a master 
equation, the disorder-averaged master equation has a memory 
effect \cite{Klafter80}. Thus, the dynamics of an averaged quantity may be 
non-Markovian even though the process for each sample is Markovian.
There are several reports on absence of SA in transport phenomena 
in disordered media: for mean square displacement \cite{Lopez95}, for 
mean first-passage time of the Sinai model \cite{Noskowicz88}.

In this letter, we study a random walk model in 
a one dimensional disordered medium with an oscillatory input current as 
a generic model of the BPM. 
It is rigorously shown that the admittance, which is 
equal to the Fourier-Laplace transform of 
the first-passage time distribution (FPTD), 
is non-self-averaging (non-SA) when the disorder is strong.

We consider a general random walk in a one dimensional lattice segment of 
$N+1$ sites.
The lattice sites are denoted by integers, $n=0,1,\ldots,N$.
The dynamics of a random walking particle can be described by 
a master equation for the probability $P_n(t)$ that the particle is 
at site $n$ at time $t \geq 0$.
The master equation for the model is written as
\begin{equation}
\dot{P}_n(t)=w_{n \ n-1}P_{n-1}(t)-(w_{n-1 \ n}+
w_{n+1 \ n})P_{n}(t)+w_{n \ n+1}P_{n+1}(t),
\label{master1}
\end{equation}
where $w_{m \ n}$ denotes the random 
jump rate of a particle from site $n$ to site $m$.
The probability distribution of jump rates characterizes the random medium. 
We introduce a perturbation at the left end, 
so that the equation for $P_0(t)$ 
is given by 
\begin{equation}
\dot{P}_0(t)=-w_{1 \ 0}P_{0}(t)+w_{0 \ 1}P_1(t)+J(t).
\label{master2}
\end{equation}
$J(t)$ is the oscillatory current perturbation at site $0$. 
The right end of the system is assumed to be a sink and the equation for 
$P_{N-1}$ is given by
\begin{equation}
\dot{P}_{N-1}(t)=w_{N-1 \ N-2}P_{N-2}(t)-(w_{N-2 \ N-1}+w_{N \ N-1})P_{N-1}(t).
\label{master3}
\end{equation} 
Since the current perturbation $J(t)$ oscillates in time around 
a positive average with 
the amplitude $\triangle J$, the response of the output current 
$w_{N \ N-1} P_{N-1}(t)$ oscillates around its stationary state 
with the amplitude $w_{N \ N-1} \triangle P_{N-1}$ at the same frequency 
with a phase-shift. 
The admittance is defined as the ratio of these two amplitudes
\begin{equation}
\chi(\omega)\equiv \frac{w_{N \ N-1} \triangle P_{N-1}}{\triangle J}.
\end{equation}

We first note that the admittance can be related to the first passage time 
distribution $F_{N \ 0}(t)$ (FPTD), which is the probability density that 
the particle which starts at site $0$ at time $0$ 
arrives for the first time at site $N$ at time $t$.
Since site $N$ is a sink, the FPTD is given by the output current 
from site $N$ when there is no input current 
and the particle starts from site $0$ at time $0$.
The Fourier-Laplace transform of 
a system of master equations for the case is the same 
as a system of equations for $\triangle P_n / \triangle J$ derived from 
the master equations Eqs.\ (\ref{master1},\ref{master2},\ref{master3}). 
It implies that $\tilde{F}_{N \ 0}(s=i \omega)$ for the case where there is no 
input current and the particle starts from site $0$ at time $0$ 
is identical to $ w_{N \ N-1} \triangle P_{N-1}/ \triangle J$.
Thus, the admittance is equal to the Fourier-Laplace 
transform of the output current in the problem described above, i.e. the FPTD.

We make the low-frequency expansion of the admittance to see 
the behavior near the static limit.
Since the admittance is given by the Fourier-Laplace transform of the FPTD, 
the admittance at zero frequency equals to one due to normalization 
of the FPTD. 
The mean first-passage time (MFPT) $\bar{t}$ which is the first moment 
of the FPTD is given by $\bar{t}=i \ d\chi(\omega = 0)/d\omega$.
Thus, the low-frequency expansion of the admittance is given by 
\begin{equation}
\chi(\omega)=1-i \ \bar{t} \ \omega+O(\omega^2). \label{adm}
\end{equation}
Since the MFPT is given by \cite{Murthy89}
\begin{equation}
\bar{t}=\sum^{N-1}_{k=0} \frac{1}{w_{k-1 \ k}}+
\sum^{N-2}_{k=0}\frac{1}{w_{k+1 \ k}}\sum^{N-1}_{i=k+1} \prod^{i}_{j=k+1}
\frac{w_{j-1 \ j}}{w_{j+1 \ j}}, \label{mfpt}
\end{equation}
the MFPT $\bar{t}$ is larger than $\sum^{N-1}_{k=0} 1/w_{k-1 \ k}$.
By taking the average of Eq.\ (\ref{mfpt})
over the distribution for the jump rates,
one sees that the average of the MFPT 
$\langle \bar{t} \rangle$ 
is larger than the first inverse moment of a random jump rate 
$\langle 1/w \rangle$.
Thus, when the first inverse moment of a jump rate diverges, 
which is called strong disorder, the disorder-averaged MFPT diverges.
It indicates that the disorder-averaged admittance is non-analytic.
Since the admittance is a function of $i \omega$, we can write as 
\begin{equation}
\langle \chi(\omega) \rangle -1 \sim (i \omega)^{\mu} \label{achi},
\end{equation}
where $0 < \mu < 1$. 
Since this behavior of the disorder-averaged admittance is 
completely different from the behavior of 
the admittance of each sample given by Eq.\ (\ref{adm}), 
the admittance is non-SA when the disorder is strong.

On the other hand, when the first inverse moment of a jump rate exists,
it is easily shown from Eq.\ (\ref{mfpt}), that the MFPT is finite 
provided that there is no correlation between $w_{k+1 \ k}$ and 
$w_{i+1 \ i}$, where $i \geq k+1$.
The condition of independence of the neighboring jump rates holds 
for the site disordered model called random trap model and 
the bond disordered model called random barrier model.
Thus, when the first inverse moment of a jump rate exists,
the low frequency behavior of the disorder-averaged admittance is 
as same as the one of each sample given by $\chi -1 \sim i \omega$.
Furthermore, if the second inverse moment of a jump rate exists,
the prefactor of $\omega$ does not depend on a realization of the chain 
when the chain is sufficiently long.

\begin{figure}
\centerline{\epsfxsize=8cm
\epsffile{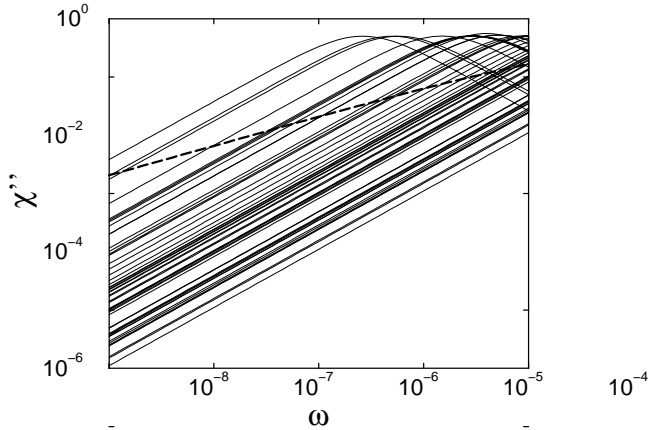}}
\caption{Low-frequency behavior of the imaginary part of the admittance 
when the exponent of the probability distribution of jump rates 
is $1/2$ and the length of a chain is $100$.
Each solid line represents the admittance for each of 50 samples.
The dashed line represents the admittance averaged over 5000 samples.
Though the admittance of each sample is proportional to $\omega$,
the averaged admittance is proportional to $\sqrt{\omega}$.}
\label{low-frequency}
\end{figure}

In order to test the foregoing observation, we performed a computer simulation
in one dimensional chain of $100$ sites.
Figure \ref{low-frequency} shows the low frequency behavior of 
the imaginary part of the admittance 
obtained by the Monte-Carlo simulation where the relation 
$w_{i+1 \ i}=w_{i \ i+1}$ is assumed and the probability distribution of 
jump rates is given by 
\begin{equation}
 P(w) = \left\{ \begin{array}{ll}
			\alpha w^{\alpha-1} & \mbox{if $0<w<1$} \\
			0                  & \mbox{otherwise,}
		  \end{array}
	  \right.  \label{jrdist}
\end{equation}
where $0 < \alpha < 1$.
The power-law probability distribution corresponds to 
the waiting-time distribution 
$\Psi (t) \sim t^{-\alpha-1}$ used in the study of dispersive transport 
where $\alpha$ is proportional to temperature \cite{Pfister78}, 
and it has been shown that the power-law distribution 
is common for activation processes with random activation energy 
\cite{Odagaki95}.

In the literature of experiments of the BPM \cite{Yasuda95,Jongh96}, 
the Cole-Cole plot of the admittance is employed for the analysis.
It is a parametric plot of the imaginary part of the admittance 
against the real part. 
The shape of the Cole-Cole plot is independent of the length 
of the chain for a sufficiently long chain due to 
time-scale invariance of the Cole-Cole plot.
Thus, the Cole-Cole plot shows the size-independent characteristics 
of a medium.

In order to see the non-SA property of the admittance in the Cole-Cole plot,
we first note that the inverse of the admittance satisfies the 
following recursive relation:
\begin{equation}
\chi_{n}^{-1}(\omega)=(\frac{i \omega}{w_{n+1 \ n}}+1+
\frac{w_{n-1 \ n}}{w_{n+1 \ n}}) \chi^{-1}_{n-1}(\omega)-
\frac{w_{n-1 \ n}}{w_{n+1 \ n}} \chi^{-1}_{n-2}(\omega),
\label{rec}
\end{equation}
where $\chi_{n}^{-1}(\omega)$ is the inverse of the admittance 
for a chain of length $n$ \cite{unpublished}.
The recursive relation proves inductively 
that the inverse of the admittance obeys 
the following four inequalities in the frequency region $[0, \omega_n]$, 
where $\omega_n$ is the positive smallest zero of ${\chi_n^{-1}}'$:
\begin{equation}
\left\{ \begin{array}{l}     {\chi_n^{-1}}'' > 0 \\
	                     {\chi_{n-1}^{-1}}' > 0 \\
	                     {\chi_n^{-1}}'' > {\chi_{n-1}^{-1}}'' \\
	                     {\chi_n^{-1}}' < {\chi_{n-1}^{-1}}'.
	\end{array} 
\right.
\end{equation}
Since the second inequality shows $\omega_n \leq \omega_{n-1}$,
the fourth inequality implies 
${\chi_n^{-1}}' < {\chi_{n-1}^{-1}}' < \cdots < {\chi_1^{-1}}'=1$ 
in the region $[0,\omega_n]$. It is straightforward to show that 
${\chi^{-1}}' < 1$ implies that the Cole-Cole plot of the 
admittance of each sample can exist only 
{\em outside a semicircle} whose center is at $(1/2,0)$ and 
radius is $1/2$, 
which we call the Debye semicircle, when $\omega < \omega_n$.

\begin{figure}
\centerline{\epsfxsize=8cm
\epsffile{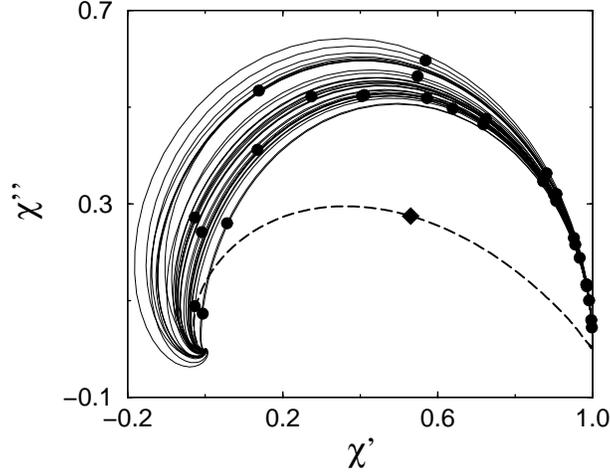}}
\caption{The Cole-Cole plot of the admittance 
when the exponent of the probability distribution of jump rates is $1/2$ and
$N=100$. Each solid line represents the admittance for each of 30 samples.
The dashed line represents the admittance averaged over 5000 samples.
The circles represents the admittance of each sample 
at $\omega = 5 \times 10^{-7}$.
Although the circles scatter outside the Debye semicircle, 
the diamond, i.e. the averaged admittance is inside of Debye semicircle.}
\label{Cole-Cole}
\end{figure}

On the other hand, we can show in the following way that the
Cole-Cole plot of the averaged admittance is located  
{\em inside the Debye semicircle} 
when the disorder is strong.
We consider the angle $\theta$ between the tangent of the Cole-Cole plot at 
(1, 0) and the horizontal axis. For Eq.\ (\ref{achi}), $\theta=\mu \pi/2$, 
which is less than $\pi/2$ since $\mu < 1$, is obtained. 
However the angle $\theta$ for the Debye semicircle is equal to $\pi/2$.
It implies that the Cole-Cole plot of 
the averaged admittance appears inside the Debye semicircle 
when the disorder is strong. 
Since the Cole-Cole plot of the admittance of any sample
appears outside the Debye semicircle, 
the Cole-Cole plot clearly shows non-SA property of the admittance.
Figure \ref{Cole-Cole} shows the Cole-Cole plot of $ \chi $ and 
$\langle \chi \rangle$ obtained by Monte Carlo simulation
when $\alpha=1/2$ in Eq.\ (\ref{jrdist}).
Since the admittance of each sample at the same frequency scatters outside the
Debye semicircle, the admittance averaged at the same frequency 
is inside the Debye semicircle.

In the case that the first inverse moment of a jump rate exists,
the angle $\theta$ of the Cole-Cole plot of the averaged admittance 
is $\pi/2$, which is as same as the Cole-Cole plot of each sample, 
because of the regular behavior as $\langle \chi \rangle -1 \sim i \omega$.

In conclusion, we have presented a generic stochastic dynamical model in 
one dimensional medium of the BPM and have shown that the admittance 
has important information about transport in disordered media, e.g. the FPTD.
Thus the BPM will be a powerful technique to investigate 
the dynamical process in random media.
We have also shown that the admittance is non-SA 
when the disorder is strong. Since the absence of SA is due to 
non-analyticity introduced by an average over 
infinite number of samples,
the behavior of the averaged admittance over infinite number of samples 
given by a theoretical analysis does not coincide with that of 
the admittance observed in experiments. 

Non-SA behavior, i.e. strong sample to sample dependence 
may be observed in other experiments. For example, 
it is known that anomalous system-size dependence of the mobility, 
$\mu \sim N^{1-1/\alpha}$, is observed in the  
time-of-flight experiment of amorphous semiconductors \cite{Pfister78}.
The anomalous system-size dependence is due to the fact that 
the smallest value of random jump rates depends on the length of a chain 
\cite{Kehr98}. Since the smallest value 
dominates the mobility, the mobility is also a random variable.
Thus, the anomalous system-size dependence implies absence of SA.

It is important to note that the system treated in the present letter 
is purely one-dimensional and it is still an open problem 
whether the admittance for higher dimensional media are SA.
Since one can consider a diffusing particle's path a one dimensional chain 
and a higher dimensional medium can be regarded as a bundle of 
many realizations of a particle's path,
the total output from the medium is given by the sum of one dimensional 
currents. Thus, the admittance for such a system may be SA,
since the number of paths is very large when the medium is macroscopic.
Thus, in order to see the non-SA behavior described here 
one needs to work in one dimensional system or the case where 
the translational invariance is violated in only one direction 
since there is still possibility that the SA assumption is valid for 
two or three dimensional systems.

This work was supported in part by Grant-in-Aid from 
the Ministry of Education, Science, Sports and Culture.
One of us (M.K.) is thankful to the Institut 
f\"{u}r Festk\"{o}rperforschung, 
Forschungszentrum J\"{u}lich for its hospitality,
where part of the present work was done.



\begin{references}
\bibitem{Yasuda95}Y. Yasuda, K. Iwai, and K. Takakura, J. Phys. Chem. 
{\bf 99}, 17852 (1995); Y. Yasuda, Heterogen. Chem. Rev. {\bf 1}, 103 (1994).
\bibitem{Jongh96}P. E. de Jongh and D. Vanmaekelbergh, Phys. Rev. Lett. 
{\bf 77}, 3427 (1996); G. Franco, J. Gehring, L. M. Peter, E. A. Ponomarev, 
and I. Uhlendorf, J. Phys. Chem. B {\bf 103}, 692 (1999).
\bibitem{general interest}S. Alexander, J. Bernasconi, W. R. Schneider, 
and R.Orbach, Rev. Mod. Phys. {\bf 53}, 175 (1981);
J. W. Haus and K. W. Kehr, Phys. Rep. {\bf 150}, 263 (1987);
G. H. Weiss and R. J. Rubin, Adv. Chem. Phys. {\bf 52}, 363 (1983);
J. P. Bouchaud and A. Georges, Phys. Rep. {\bf 195}, 127 (1990).
\bibitem{Klafter80}J. Klafter and R. Silbey, Phys. Rev. Lett. {\bf 44}, 
55 (1980).
\bibitem{Lopez95}J. M. L\'{o}pez, M. A. Rodr\'{\i}guez, and L. Pesquera, Phys. 
Rev. E {\bf 51}, R1637 (1995).
\bibitem{Noskowicz88}S. H. Noskowicz and I. Goldhirsch, Phys. Rev. Lett. 
{\bf 61}, 500 (1988).
\bibitem{Murthy89}K. P. N. Murthy and K. W. Kehr, Phys. Rev. A {\bf 40}, 
2082 (1989).
\bibitem{Pfister78}G. Pfister and H. Scher, Adv. Phys. {\bf 27}, 747 (1978).
\bibitem{Odagaki95}T. Odagaki, Phys. Rev. Lett. {\bf 75}, 3701 (1995).
\bibitem{unpublished}The recursive Eq. (\ref{rec}) is derived from the
following two equations for the FPTD obtained by its definition:
$\tilde{F}_{i+1 \ i}(s)=\tilde{\psi}_{i+1 \ i}(s)+\tilde{\psi}_{i-1 \ i}(s)
\tilde{F}_{i+1 \ i-1}(s)$ and 
$\tilde{F}_{i+1 \ i-1}(s)=\tilde{F}_{i+1 \ i}(s) \tilde{F}_{i \ i-1}(s)$,
where $\tilde{F}_{i \ j}(s)$ denotes the Laplace transform of the FTPD
from site $j$ to site $i$ and $\tilde{\psi}_{i \ j}(s)$ denotes the Laplace 
transform of the waiting time distribution for the jump 
from site $j$ to site $i$.
\bibitem{Kehr98}K. W. Kehr, K. P. N. Murthy, and H. Ambye, Physica A 
{\bf 253}, 9 (1998).
\end{references}
\end{document}